\documentclass[conference]{IEEEtran}
\IEEEoverridecommandlockouts
\usepackage{cite,amsmath,amssymb,graphicx,booktabs,url,microtype}
\usepackage[hidelinks]{hyperref}
\hypersetup{
  pdftitle={Strategic Evaluation of Planning Strategies for LLM Agents in Cyber-Physical Systems},
  pdfauthor={J. de Curto and I. de Zarza}
}

\usepackage{dsfont}
\newcommand{\ind}{\mathds{1}}

\begin{document}

\title{Strategic Evaluation of Planning Strategies for LLM Agents in Cyber-Physical Systems}

\author{
\IEEEauthorblockN{J. de Curt\`o\IEEEauthorrefmark{1}\IEEEauthorrefmark{2}
                  and I. de Zarz\`a\IEEEauthorrefmark{2}\IEEEauthorrefmark{3}}
\IEEEauthorblockA{\IEEEauthorrefmark{1}\textit{Department of Computer Applications in Science and Engineering},
\textit{BARCELONA Supercomputing Center}, Barcelona, Spain}
\IEEEauthorblockA{\IEEEauthorrefmark{2}\textit{Escuela T\'ecnica Superior de Ingenier\'ia (ICAI)},
\textit{Universidad Pontificia Comillas}, Madrid, Spain\\
Email: jdecurto@icai.comillas.edu}
\IEEEauthorblockA{\IEEEauthorrefmark{3}\textit{Human-Centered AI, Data and Software},
\textit{LUXEMBOURG Institute of Science and Technology},\\
Esch-sur-Alzette, Luxembourg\\
Email: irene.zarza@list.lu}
}

\maketitle

\begin{abstract}
LLM-agent evaluations commonly measure task success or agreement with a
declared plan. In strategic cyber-physical systems, an architecture must also
remain appropriate after autonomous participants respond and physics constrains
outcomes. We introduce a controlled benchmark of \emph{planning-induced control
trajectories}: ordered planning operations and directives linking execution
architecture to strategic response and physical consequences. Four coded
executors (predefined, sequential, hierarchical, and search) control demand
response for 40 prosumers on a radial feeder. The LLM declares or advises typed
policies and mediates communication; schedules, base prosumer dynamics,
stochastic actions, and power flow remain explicit code. Paired forced-mode
counterfactuals, exact-prompt caching, common response draws with separate
randomness streams, critic isolation, and event-level feasibility isolate
comparisons. The Llama-3.3-70B experiments on this single feeder distinguish three
properties. First, forced search is the oracle in all five baseline seeds under
the specified objective. Second, injected objective substitution preserves mode
agreement at 1.0 while increasing cumulative voltage shortfall by $2.68\times$.
Third, the 144-scenario, 576-episode factorial bank, using three repeated seeds,
contains feasible oracles from predefined, sequential, and search. The
prespecified stress-held-out ridge has mean regret 90.7 and no observed value
over fixed sequential. A secondary post-hoc constraint-aware analysis reduces
regret to 29.0; a simple deadline rule attains 28.7, so this gain does not establish
a learning advantage. An all-feasible ablation does not improve over fixed
search. These are simulation-internal, descriptive comparisons. A five-model,
300-declaration extension tests interface behaviour, not cross-backbone physical
rankings; shared-endpoint latency tails motivate probabilistic live feasibility.
\end{abstract}


\section{Introduction}

Planning is becoming an explicit systems component of LLM agents. Rather than
relying on one undifferentiated reasoning loop, an agent may construct a plan
once, revise it sequentially, decompose it hierarchically, or search over
candidate trajectories~\cite{yao2023react,wang2023plan,yao2023tree,shinn2023reflexion}.
Making this choice explicit improves auditability: Oota et al.~\cite{oota2026llmagentsexecuteplans}
separate planning-mode selection from execution through \emph{Planning-as-Routing},
dispatching declarations to pattern-specific executors.

In strategic cyber-physical systems, this separation must also account for
autonomous response and physical constraints. Agreement between declared and
executed modes does not show
that the selected strategy was appropriate, that its subgoals were grounded in
the available state, that autonomous participants would comply, or that the
resulting physical trajectory was safe. Conventional agent benchmarks compress
different decompositions, revisions, retries, and recovery paths into terminal
scores~\cite{deng2023mind2web,zhou2024webarena,jimenez2024swebench,shridhar2021alfworld}.
Trace-level analysis recovers some of this structure, but an evaluator expressed
in the same language representations as the agent remains partly
self-referential.

We therefore separate three evaluation dimensions. \emph{Planning-strategy
heterogeneity} asks whether operationally distinct architectures induce
different trajectories and outcomes under matched conditions. \emph{Execution
fidelity} asks whether the realised trajectory follows both the declared
architecture and the intended objective. \emph{Adaptive selection} asks whether
pre-decision state can identify a low-cost, feasible architecture. These
distinctions are acute in multi-agent cyber-physical systems: planning decisions
alter autonomous participants, whose realised actions then drive physics. We call
the ordered commitments, decompositions, revisions, candidate evaluations, and
directives a \emph{planning-induced control trajectory}. It is the decision-layer
cause, not the electrical or mechanical state path that emerges after response.

Demand response in smart grids provides a rigorous testbed for this distinction. A distribution
operator requests flexibility from heterogeneous prosumers while respecting
feeder-head import and voltage constraints. The system combines strategic
private costs, a natural spatial hierarchy for decomposition, and an external
power-flow model that can determine whether a plausible-looking directive was
physically useful. The domain is therefore an instrument for studying planning,
not merely an application of an LLM to a grid.

The operational task is concrete. Whenever forecast feeder import exceeds the
1.1~MW cap or voltage approaches the 0.95~pu floor, the operator must decide
how to organise a curtailment campaign: which execution architecture to use,
which nodes to target, how strongly to request flexibility, and how to frame the
request. The objective in Eq.~\eqref{eq:objective} rewards removal of cap and
voltage violations while penalising unnecessary curtailment. In the uncontrolled
calibration trace, for example, hour 17 reaches 1.644~MW at the feeder head and a
minimum voltage of 0.9344~pu. The controller must therefore coordinate actual
flexible demand, not merely produce a plausible verbal plan.

The LLM's control boundary is deliberately narrow. It does not solve power
flow, construct the numerical node allocation, or decide a household's action.
It operates at two interfaces: high-level policy declaration or bounded mode
advice, and language-mediated communication. Typed strategy executors convert
policy fields into schedules; a game-theoretic response model, augmented by a
bounded LLM persuasion shift, determines prosumer behaviour; and DistFlow
independently evaluates the physical consequences. We use \emph{LLM agent} for
this composite controller rather than for an unconstrained language model acting
directly on the feeder. Forced-mode comparisons therefore evaluate coded
executors within this controller, not autonomous LLM plan synthesis.

Accordingly, we study three questions. \emph{RQ1:} do distinct planning
architectures induce measurably different strategic and physical outcomes under
matched operating conditions? \emph{RQ2:} does faithful execution of a declared
mode also preserve the intended physical objective? \emph{RQ3:} can a selector
exploit scenario-dependent strategy heterogeneity under observability and
real-time constraints? We answer through a four-layer pipeline comprising
strategy selection, mode-specific execution, strategic response, and independent
physical verification. Every forced comparison keeps the objective, agents,
forecast, latent response draws, and network dynamics fixed; paired
per-scenario regret then measures selection quality directly.

\textbf{Contributions.} (1) We introduce a controlled, physics-grounded
benchmark that isolates planning architecture from strategic adaptation and
physical evolution. (2) We define planning-induced control trajectories as a
common abstraction for comparing predefined, sequential, hierarchical, and
search execution independently of the application domain. (3) We develop a
counterfactual protocol combining paired forced execution, exact-prompt caching,
common random numbers, event-level feasibility, response-aware metrics, and
external physical verification. (4) On the tested bank, scenario-dependent oracle diversity defines a
selection problem, while known feasibility explains the constraint-aware routing
gain. (5) A five-model interface extension shows that declaration collapse is
model-dependent and that serving latency has long observed tails, motivating
risk-aware rather than purely deterministic deployment gates. Code and data have been released in the GitHub repository. \footnote{\url{https://github.com/drdezarza/LLMstrategicplanning}}

\section{Related Work}

\textbf{Planning strategies for LLM agents.} ReAct interleaves reasoning and
action~\cite{yao2023react}; Plan-and-Solve makes decomposition explicit
\cite{wang2023plan}; Tree of Thoughts searches over candidate reasoning paths
\cite{yao2023tree}; and Reflexion introduces verbal feedback across attempts
\cite{shinn2023reflexion}. These approaches differ in execution structure,
compute, observability, and recovery, not merely in prompt wording. Agent
benchmarks make these differences visible at the task level
\cite{deng2023mind2web,zhou2024webarena,jimenez2024swebench,shridhar2021alfworld},
while broader evaluation work argues for multidimensional rather than scalar
assessment~\cite{liang2022holistic}. Oota et al.~\cite{oota2026llmagentsexecuteplans}
study the declaration--execution gap and mode selection across web, coding, and
embodied tasks. We complement this with strategic response, physical-objective
fidelity, and event-level feasibility.

\textbf{Faithfulness and trajectory evaluation.} Chain-of-thought studies show
that stated rationales can diverge from the effective basis of an answer
\cite{lanham2023measuring,turpin2023language}. The analogous plan-level gap is
that a verifier may certify the named executor while remaining blind to target
substitution, unsupported state assumptions, or physically ineffective actions.
We therefore combine trace-derived quantities with an external physical
referent.

\textbf{Strategic multi-agent systems.} LLM-agent populations have been studied
as social simulations and strategic systems
\cite{park2023generative,guo2024large,xi2023rise}. Controlled experiments examine
finitely repeated games~\cite{akata2025playing} and common-pool-resource
dilemmas~\cite{piatti2024cooperate}. Our response model builds on cooperation
theory~\cite{axelrod1984evolution,nowak2006five} and prior work on coevolutionary
adaptation and bounded LLM influence
\cite{deCurto2025influence}. The LLM does
not directly set prosumer actions; it provides a bounded modulation of a
heterogeneous game-theoretic base probability.

\textbf{Cyber-physical evaluation and smart grids.} Demand response coordinates
flexible consumption under network and market constraints
\cite{siano2014demand,parag2016electricity}. AI methods are increasingly used in
demand response~\cite{khan2022artificial}, and recent work examines LLM
integration in smart grids~\cite{madani2025large,shi2024review}. Here the grid is
methodological: linearised DistFlow~\cite{baran1989network}, a standard radial
model and basis of later branch-flow relaxations~\cite{farivar2013branch},
provides an evaluator independent of the planner.

The demand-response environment used here extends the microgrid coordination setting of \cite{deCurto2026}, which modelled prosumer compliance as a repeated Prisoner's Dilemma on a social network driven by an LLM influence compiler; the present work replaces that influence compiler with a plan-declaring planner and adds network physics as an external referent for plan quality.

\section{Strategic Evaluation Framework}

\begin{figure*}[t]
\centering
\includegraphics[width=0.96\textwidth]{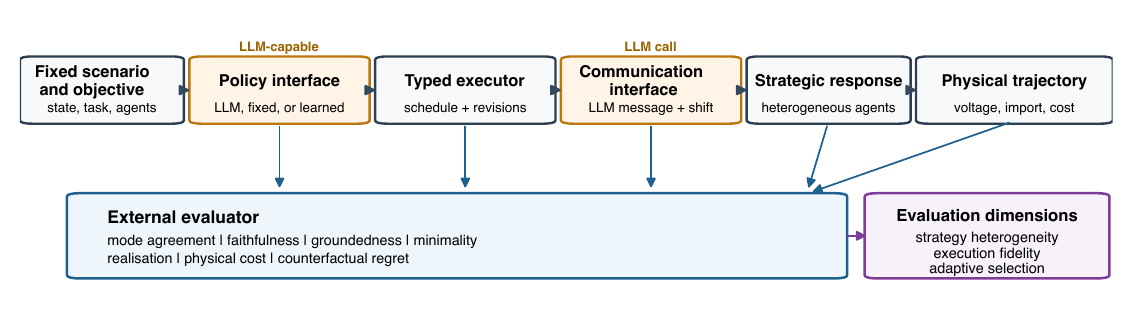}
\caption{Strategic evaluation and isolation logic. Orange boxes mark the two
bounded LLM interfaces: policy declaration/advice and language-mediated
communication. In forced counterfactuals, the policy interface is replaced by a
fixed typed policy. Strategy executors, the prosumer base game, stochastic
actions, and power flow remain explicit code; the external evaluator resolves
planning-strategy heterogeneity, execution fidelity, and adaptive selection.}
\label{fgr:framework}
\end{figure*}

Let $s_t$ denote the observed system state and $g_t$ the control objective. The
strategy set is
$\mathcal{M}=\{\textsc{predefined},\textsc{sequential},
\textsc{hierarchical},\textsc{search}\}$. A selector $\rho$ chooses
$m_t\in\mathcal{M}$, and the corresponding executor $E_{m_t}$ emits a structured
policy $p_t$. Strategic agent $o$ responds according to $B_o$, and the physical
system evolves through $F$:
\begin{align}
 m_t &= \rho(s_t,g_t), & p_t &= E_{m_t}(s_t,g_t), \\
 a_t^o &\sim B_o(p_t,h_t^o), & s_{t+1} &= F(s_t,\mathbf{a}_t),
\end{align}
where $h_t^o$ is local interaction history. Over an episode, strategy $m$
generates
\begin{equation}
\xi_m=\bigl((z_t,u_t)\bigr)_{t=0}^{T-1},
\label{eq:trajectory}
\end{equation}
where $z_t$ records planning operations and $u_t$ contains the issued control
directives. Figure~\ref{fgr:framework} separates this decision-layer trajectory
from the strategic and physical trajectories it induces. Matched forced modes
measure planning-strategy heterogeneity; trace and physical metrics measure
execution fidelity; and paired oracle regret measures adaptive selection.

\textbf{LLM interfaces and control boundary.} Table~\ref{t:llminterfaces}
shows every language-model call site in the reported system. The policy
interfaces receive a compact, serialised state rather than raw feeder arrays or
power-flow equations. Their outputs are parsed as JSON, clipped to admissible
ranges, and statically type-checked before an executor can act. No free-form text
is translated directly into kW commands. The communication interfaces are also
bounded: the LLM writes the operator message and estimates only its incremental
persuasive effect; the prosumer's base utility, memory, resistance, random action,
and realised curtailment remain outside the model.

\begin{table*}[t]
\centering
\footnotesize
\setlength{\tabcolsep}{3.5pt}
\renewcommand{\arraystretch}{1.08}
\caption{LLM interfaces in the live protocol. Strategy executors, the
base prosumer game, stochastic actions, and feeder physics are explicit code.}
\label{t:llminterfaces}
\begin{tabular}{p{0.13\textwidth}p{0.25\textwidth}p{0.25\textwidth}p{0.29\textwidth}}
\toprule
Interface & Prompt payload & Constrained output & Use in the reported experiments \\
\midrule
Natural policy declaration & Hour; stress and burden; remaining horizon;
volatility and spatial spread; feeder cap and forecast head import; previous
minimum voltage; recent compliance. & Typed JSON containing mode, target rule,
message theme, hierarchy depth, search branching, replanning trigger, intensity,
confidence, and a one-sentence rationale. & E1 natural declaration and the
natural routed arm of E2. The interface is replaced by fixed policies in forced
counterfactual banks. \\
Hybrid mode advice & The same state plus deterministic feature-derived base
scores for all four modes. & A per-mode shift in $[-0.30,0.30]$ and a short
reason. Code applies the shifts, samples the mode, and constructs the remaining
policy fields. & E1/E2 bounded hybrid and E4 before the deterministic deadline
gate. The LLM does not directly choose the final mode. \\
Operator message generation & Message framing, dispatch hour, and forecast kW
above the import cap. & A specific two--three sentence customer-facing
curtailment request. & Every live episode with a non-empty dispatch; one message
is reused for all targets in that hour. \\
Prosumer message evaluation & Archetype description, requested kW, neighbour
compliance, exploitation count, recent response history, and the operator
message. & A willingness shift in $[-0.30,0.30]$, a persuasion flag, and one short
reason. The shift is resistance-attenuated and cannot itself choose the action. &
Applied to a common-random-number sample of 40\% of targeted prosumers; E6
attributes the resulting bounded shifts and actions. \\
Optional plan critic & Proposed typed policy, state summary, decision deadline,
and stipulated executor cost. & Approval, risk level, and adjusted intensity. &
Implemented for ablation, but the LLM critic is disabled in the reported causal
banks. E4 uses only the deterministic, type-valid deadline gate. \\
\bottomrule
\end{tabular}
\end{table*}

\textbf{Illustrative dispatch.} At hour 17 of the uncontrolled calibration
trace, feeder import is 544~kW above the cap and minimum voltage is 0.9344~pu.
The planner is not asked to produce appliance commands. It receives the state
summary and may emit a type-valid policy such as
\{\texttt{mode=SEARCH}, \texttt{target=HIGH\_LOAD},
\texttt{branching=4}, \texttt{replan=NONE}\}. The search executor then
constructs targeting--intensity schedules, evaluates each through forecast
DistFlow, and retains the lowest predicted-cost schedule. For the selected
households, the message prompt receives the framing, hour, and kW exceedance; a
second prompt receives each sampled prosumer's personality, requested reduction,
neighbour behaviour, and recent history, and returns only a bounded willingness
shift. The game-theoretic probability plus that shift determines stochastic
compliance, after which the physical simulator, not the LLM, computes import,
voltage, and cost. This end-to-end example is the intended meaning of an LLM
planning agent in the benchmark.

\textbf{Typed policy and execution semantics.} A policy records
\texttt{mode}, targeting rule, hierarchy depth, search branching, replanning
trigger, intensity, message theme, and decision budget. Static checks detect, for
example, a hierarchical policy without decomposition or search with fewer than
two candidates. All four executors receive the same implemented objective: remove
feeder-head excess above 1.1~MW and voltage shortfall below 0.95~pu while
limiting unnecessary curtailment, as formalised in Eq.~\eqref{eq:objective}.
\textsc{Predefined} compiles the remaining-horizon allocation once and commits.
\textsc{Sequential} dispatches a step, observes realised response, and
reconstructs the unexecuted suffix. \textsc{Hierarchical} allocates the required
reduction first across laterals and, at depth three, feeder segments.
\textsc{Search} generates complete targeting--intensity schedules, rolls them out
through forecast power flow, and retains the lowest predicted-cost candidate.
Thus the experiment varies when the controller commits, whether it revises or
decomposes, and whether alternatives are explicitly evaluated.

\textbf{Strategic response.} Each of 40 prosumers in a smart grid has an archetype, resistance,
flexible load, neighbourhood, and memory. Compliance starts from a
cooperation-based probability combining reciprocity, exploitation memory,
comfort fatigue, and archetype bias. For 40\% of targeted agents, an LLM evaluates
the operator narrative and returns a bounded shift
$\delta\in[-0.30,0.30]$, attenuated by resistance. Compliance and realised
curtailment remain stochastic agent decisions rather than direct LLM outputs.

\textbf{External evaluation.} The executed-mode verifier checks architecture
agreement and policy type. The external evaluator additionally measures:
faithfulness (Jaccard overlap of declared and issued targets), groundedness (share
of directives within 105\% of true flexible load), minimality (oracle divided by
realised curtailment on successful hours), coherence (one minus normalised
revision churn), and directive realisation (realised divided by requested power).
The physical objective is
\begin{equation}
J=\mathrm{kWh}_{\mathrm{overcap}}+4000\sum\Delta v
  +0.25\,\mathrm{kWh}_{\mathrm{curtailed}},
\label{eq:objective}
\end{equation}
where lower is better. The coefficients encode benchmark priorities, not
calibrated economic costs. For scenario $q$, paired routing regret is
\begin{equation}
R_\rho(q)=J\!\left(\rho(q),q\right)-\min_{m\in\mathcal{M}}J(m,q).
\label{eq:regret}
\end{equation}

\section{Experimental Methodology}

\textbf{Environment and model.} The 24-hour feeder has four laterals of ten
prosumer nodes ($N=40$), $S_{\mathrm{base}}=1$~MVA, per-segment
$r=0.022$ and $x=0.014$ pu, a 1.1~MW import cap, and a 0.95 pu voltage
floor. Cooling-dominated load and declining afternoon PV produce eight baseline
violation hours. The reported live run uses
Llama-3.3-70B-Instruct~\cite{meta2024llama33}, a 70B instruction-tuned member of
the Llama 3 family~\cite{grattafiori2024llama3}, with temperature 0.2 and at most
400 completion tokens. E1, E2, and E5 use seeds
$\{7,13,42,101,202\}$; E3, E4, E6, and each E7 factor combination use
$\{7,13,42\}$.

\textbf{Isolation controls.} In paired forced-mode comparisons, all four
strategies receive the same feeder, load--PV realisation, observed forecast,
prosumer population, objective, targeting rule, narrative theme, and
stress-calibrated intensity. Forecast, planning, tool-failure, adversarial, and
response randomness use separate deterministic streams; latent narrative
sampling, compliance, and realisation draws are shared across strategies by
$(\mathrm{seed},t,\mathrm{node})$. Every cache entry is keyed by endpoint/model,
system message, and complete prompt, with a collision guard. The last scheduled
step is executed rather than replaced by a final-hour redeclaration. The LLM
critic is disabled in causal forced banks; E4 activates only a deterministic,
type-valid deadline gate. Every declaration or revision incurs a stipulated
nominal cost, and feasibility requires that no individual planning event exceed
the decision deadline.

\textbf{LLM role by experiment.} E1 and the routed arms of E2 exercise the
policy interfaces in Table~\ref{t:llminterfaces}: natural declaration asks the
LLM for the full typed policy, whereas the hybrid asks only for bounded score
shifts. The forced arms of E2, all of E3, E5, E6, and the E7 counterfactual bank
fix the mode and shared policy fields to isolate execution semantics; the LLM
continues only in message generation and bounded prosumer interpretation. E4
uses hybrid declaration followed by the deterministic deadline gate, never the
LLM critic. The E7 ridge and constraint-aware diagnostic are offline statistical
selectors over committed counterfactual data and make no additional LLM calls.

\textbf{Multimodel interface extension.} To distinguish framework effects
from properties of the Llama backbone used in E1--E7, M0--M5 replay the same
typed declaration, bounded mode-advice, and prosumer-interpretation interfaces
with Llama-3.3-70B-Instruct, DeepSeek-V4-Pro~\cite{deepseek2026v4},
Gemma-3-27B-IT~\cite{gemma2025gemma3}, GLM-5.2~\cite{zai2026glm52}, and
MiniMax-M3~\cite{minimax2026m3}. M1--M2 use 20 deterministic planner-visible
states (four stress multipliers by five decision hours) and three declarations
per state: 60 per model and 300 total. M3 records 40 bounded score-shift calls
per model; M4 records 48 archetype--message evaluations per model. M2 retains a
conservative permutation test over all 20 states and adds a secondary,
power-oriented three-band association and logistic stress trend. M5 reports
declaration-only wall time observed through one shared endpoint, separately from
the other interfaces. The extension makes no feeder rollouts and therefore tests
interface behaviour and deployment assumptions, not cross-model replication of
the E2--E7 physical rankings.

\textbf{Experiments.} E1--E4 characterize strategy use, paired physical
performance, observability, and deadline feasibility. Specifically, E1 compares
natural and bounded feature-conditioned declaration; E2 forces all four
strategies and both selectors on the same five seeds; E3 varies forecast noise
$\sigma\in\{0,.05,.10,.20,.30,.40\}$; and E4 crosses deadlines
$\{2,4,8,30\}$~s with the deadline gate. Deadlines bracket nominal executor
costs; the $\pm0.30$ shift caps bound LLM influence. Neither is calibrated to
operational data. Nominal costs exclude LLM serving latency, measured separately
in M5. E5--E6 evaluate execution fidelity and
strategic response: E5 varies aggregator honesty from 1 to 0 with the sequential
executor fixed, while E6 retains every directive for attribution by prosumer
archetype and electrical position.

E7 asks whether strategy usefulness can be learned beyond the baseline. It
crosses four feeder-stress groups, three noise levels, two deadlines, two mean
resistance levels, and three seeds: 144 scenarios and 576 forced episodes. A mode
is feasible iff it has no event-level deadline miss, and the prespecified
selection cost is
\begin{equation}
C_m(q)=J(m,q)+10^{4}\,\ind\!\left[\,m\ \text{infeasible in}\ q\,\right].
\label{eq:selectioncost}
\end{equation}
One ridge model per mode (fixed $\alpha=10$) predicts $C_m$ from cap tightness,
line resistance, forecast noise, deadline, and mean resistance. Cross-fitting
holds out an entire stress group; 5000 paired scenario-bootstrap draws yield
descriptive intervals, not independent-seed uncertainty: 48 factor combinations
reuse only three seeds. A secondary post-hoc analysis keeps the same bank,
features, folds, regularisation, and bootstrap, filters deadline-infeasible modes,
and predicts $J$ on feasible rows. A 72-scenario all-feasible subset tests quality
selection. All diagnostics use saved episodes without new model calls. Reported
$p$-values are exploratory and unadjusted unless marked $p_{\mathrm{Holm}}$; the
post-hoc Holm check covers 17 finite
tests (four E2/E5 contrasts and 13 M2 tests).
M0--M5 provide the complementary multimodel interface study: capability
probing, declaration diversity, state dependence, bounded mode advice,
persuasion structure, and observed declaration latency.

\section{Results: Three Evaluation Dimensions}

\begin{figure}[t]
\centering
\includegraphics[width=\columnwidth]{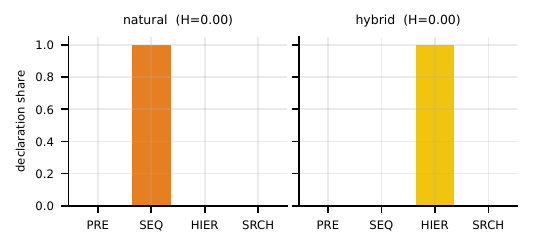}
\caption{E1 declaration shares. Natural declaration selects
\textsc{sequential} in 5/5 episodes; the bounded hybrid selects
\textsc{hierarchical} in 5/5. Both have normalised entropy $H=0$.}
\label{fgr:e1}
\end{figure}

\subsection{Declaration concentration motivates counterfactual evaluation}
Natural declaration selects \textsc{sequential} in every episode, whereas the
bounded hybrid selects \textsc{hierarchical} in every episode
(Figure~\ref{fgr:e1}); both have $H=0$. The feature prior changes which mode is
exposed, but three architectures remain unobserved under each selector. This
concentration motivates forced counterfactual evaluation of the full strategy
set. The multimodel extension below shows that this collapse is a property of
particular declarers rather than a universal consequence of the interface.

\subsection{Model choice separates three declaration regimes}

\begin{figure}[t]
\centering
\includegraphics[width=\columnwidth]{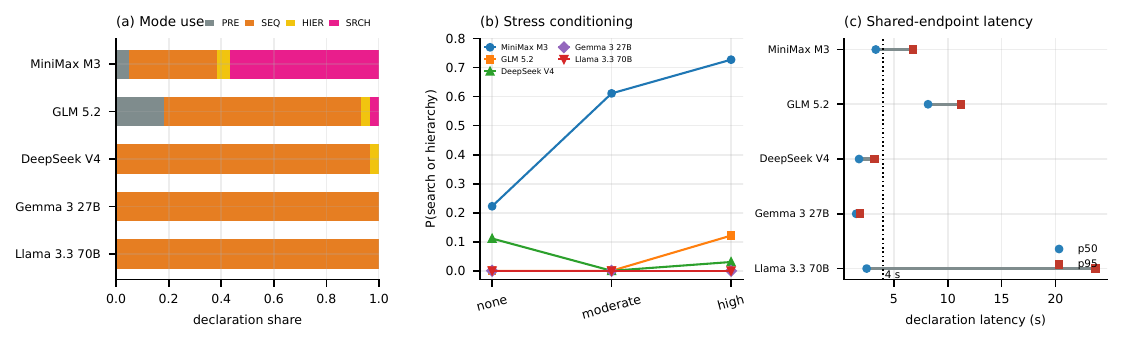}
\caption{Declaration behaviour across five models on identical planner inputs
(60 declarations per model). (a) Probability of selecting an expensive strategy
(search or hierarchy) by feeder-stress band: MiniMax rises from 0.222 to 0.727
across bands, GLM rises weakly, and DeepSeek, Gemma and Llama stay flat.
(b) Declaration latency on a shared endpoint, $p50$ (circle) to $p95$ (square),
against the 4~s event budget of Sec.~\ref{sec:deadline}; tail risks matter for
feasibility. Panel (a) separates stress-conditioned from state-blind declarers;
panel (b) shows that the deadline constraint binds at the serving layer as well
as the strategy layer.}
\label{fgr:multimodel}
\end{figure}

Across 300 declarations (60/model), every output parses, but the distributions
separate sharply. MiniMax uses all four modes ($H=0.712$ [0.550, 0.815]); from
the zero- to high-stress band, its expensive-mode share rises from 0.222 to
0.727 and its search share from 0.111 to 0.667 ($V=0.442$, $p=0.0023$;
trend $\beta=2.803$, $p=0.0129$; Figure~\ref{fgr:multimodel}a). GLM also uses all four modes
($H=0.544$) and shows nominal band association ($V=0.345$, $p=0.0343$), although
its trend is inconclusive ($p=0.062$). The conservative 20-state permutation tests give
$p=0.076$ and $0.111$, respectively, so the banded analysis is secondary.
Only MiniMax's band association survives the 17-test Holm check
($p_{\mathrm{Holm}}=0.034$). DeepSeek uses only two modes (96.7\% sequential) without measurable state
dependence; Gemma and Llama are fully invariant. Declaration diversity and evidence of stress dependence are therefore
model-dependent.

The other bounded interfaces show partial qualitative stability. DeepSeek
returns an all-zero score shift in 39/40 M3 calls, effectively reverting the
hybrid to its deterministic prior, whereas the other models use mean $\ell_1$
budgets of 0.19--0.38 without saturation. In M4, four of five models rank
idealists first and pragmatists second; all place pragmatists above opportunists,
and four show a positive descriptive response to neighbour compliance. These are
interface-level results, not cross-model replications of E2--E7 feeder outcomes.

Observed declaration latency also challenges deterministic deployment costs (Figure~\ref{fgr:multimodel}b).
Llama has shared-endpoint $p50=2.46$~s but $p95=23.68$~s, a $9.6\times$ tail;
GLM and MiniMax also exceed 4~s at $p95$, while Gemma and DeepSeek remain below
it in this run. Because these measurements mix generation, serving load, and
queueing, they are not intrinsic model speeds; they motivate
$\Pr(L_m\!\leq\!d\mid x)$ or a latency-quantile margin for live routing. 

\begin{figure}[t]
\centering
\includegraphics[width=\columnwidth]{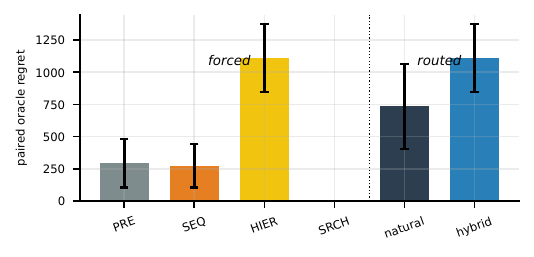}
\caption{E2 mean paired oracle regret with 95\% Student-$t$ intervals over five
seeds. Search is the per-seed oracle; routed policies provide the selection
baseline.}
\label{fgr:e2}
\end{figure}

\begin{table}[t]
\centering\small
\caption{E2 comparison over five paired seeds. Regret is computed against the
best forced strategy for the same seed. Bounded hybrid matches forced hierarchy
in these runs.}
\label{t:e2}
\begin{tabular}{lrrrr}
\toprule
Policy & $J$ & Regret & Faith. & Real. \\
\midrule
Forced \textsc{predefined}   & 1571 & 294  & 1.00 & 0.540 \\
Forced \textsc{sequential}   & 1550 & 272  & 1.00 & 0.542 \\
Forced \textsc{hierarchical} & 2388 & 1111 & 0.89 & 0.548 \\
Forced \textsc{search}       & \textbf{1277} & \textbf{0} & 0.79 & 0.533 \\
\midrule
Routed, natural              & 2012 & 734  & 1.00 & 0.486 \\
Routed, bounded hybrid       & 2388 & 1111 & 0.89 & 0.548 \\
\bottomrule
\end{tabular}
\end{table}

E2 establishes that executor choice is consequential: forced search is the
oracle in all five baseline seeds (Table~\ref{t:e2}). The bounded selector is
an additional 376.36 objective units relative to natural declaration (paired $t=5.514$,
two-sided $p=0.0053$, $p_{\mathrm{Holm}}=0.069$) because it selects the highest-cost forced strategy. This
reinforces that the target is calibrated strategy selection rather than
declaration diversity alone.

Natural routing exposes a second error channel. It always declares
\textsc{sequential}, yet incurs 462.05 additional objective units relative to forced sequential
($p=0.0019$, $p_{\mathrm{Holm}}=0.031$). Since the executor architecture is the same, the gap is caused by
other LLM-set fields such as target rule, intensity, and narrative theme. Mode
selection and within-mode policy parameterisation must therefore be evaluated
separately. Search's lower target-set faithfulness (0.79) is not evidence of
failure: legitimate candidate exploration changes the final target set, while
the physical objective verifies that the resulting trajectory is superior.

\begin{figure}[t]
\centering
\includegraphics[width=\columnwidth]{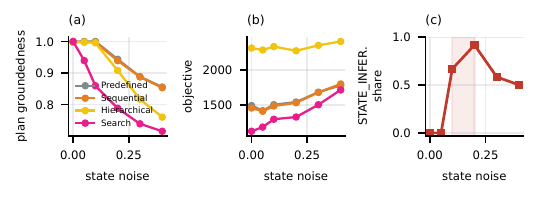}
\caption{E3 forecast-noise sweep: (a) groundedness, (b) physical objective, and
(c) share assigned the deterministic \textsc{state\_inference} failure label.}
\label{fgr:e3}
\end{figure}

Groundedness declines with noise at slopes $-0.404$ (predefined), $-0.404$
(sequential), $-0.660$ (hierarchical), and $-0.709$ (search). Objective slopes
are respectively $+821$, $+926$, $+225$, and $+1380$. Search is therefore most
sensitive on both axes, while hierarchical loses groundedness more quickly than
the flat modes without the same objective slope. The failure taxonomy assigns
\textsc{state\_inference} to 66.7\% of episodes at $\sigma=0.10$ and 91.7\% at
$\sigma=0.20$ (Figure~\ref{fgr:e3}). At larger noise, other higher-priority labels
may fire, so categorical shares need not be monotone even when forecast error
increases.

\subsection{Deadline-aware gating restores event-level feasibility}
\label{sec:deadline}
The deadline gate provides a positive feasibility control. Ungated hybrid
declaration executes hierarchical planning; at 2 and 4~s it produces nine late
planning events per episode and accumulates 51.03~s of nominal
decision cost. At 2~s the gate substitutes a type-valid sequential policy,
leaving zero misses, 13.33~s accumulated cost, and mean $J=1484.99$. At 4~s it
selects a two-candidate search policy, again with zero misses, 2.75~s accumulated
cost, and $J=1257.76$. At 8 and 30~s the gate is inert. These are mechanism and
feasibility results under stipulated per-invocation costs, not claims about
measured serving latency or the optimality of the fallback ranking.

\subsection{External physics exposes injected objective substitution}

\begin{figure}[t]
\centering
\includegraphics[width=\columnwidth]{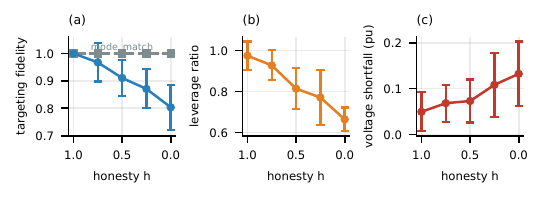}
\caption{E5 injected objective substitution. The sequential architecture continues to
match its declaration while targeting fidelity, physical leverage, and voltage
shortfall deteriorate. Error bars are 95\% Student-$t$ intervals.}
\label{fgr:e5}
\end{figure}

E5 injects an aggregator fault while retaining sequential execution: with
probability $1-h$ at each planning decision, code substitutes low-voltage-leverage
nodes and scales requested curtailment to 60\%. This is not an observed
spontaneous LLM failure. Consequently, \texttt{mode\_match}=1.0 for every row. At $h=0$, targeting
faithfulness falls to 0.804 (95\% CI [0.722, 0.885]), and physical leverage falls
from 0.975 to 0.664 (95\% CI [0.606, 0.722]; paired $p=0.00020$, $p_{\mathrm{Holm}}=0.0034$). Cumulative
voltage shortfall increases from 0.0493 to 0.1322 pu, a $2.68\times$ increase
($p=0.0041$, $p_{\mathrm{Holm}}=0.058$), while mean objective rises from 1550 to 2467. Mode
verification certifies the machinery but not the goal it pursues; the external
physical evaluator exposes the substitution directly.

\subsection{Response attribution resolves strategic heterogeneity}

\begin{figure}[t]
\centering
\includegraphics[width=\columnwidth]{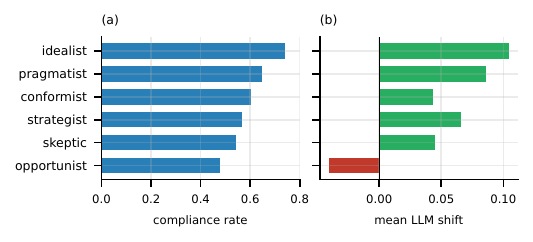}
\caption{E6 attribution over 2831 directives: (a) compliance and (b) bounded LLM
shift by prosumer archetype.}
\label{fgr:e6}
\end{figure}

Across 2831 unique directives, idealists have the highest compliance rate
(0.739) and largest mean positive LLM shift ($+0.104$). Opportunists have the
lowest compliance (0.477) and the only negative mean shift ($-0.041$). The
remaining archetypes fall between these endpoints. Directive-level
attribution therefore resolves which subpopulations support or resist a plan
instead of hiding them in aggregate realisation. Electrical position is
potentially confounded by targeting frequency and fatigue, so the experiment
supports archetype attribution but not an unconditional feeder-depth effect.

\subsection{Oracle heterogeneity creates an adaptive-selection opportunity}

\begin{figure}[t]
\centering
\includegraphics[width=\columnwidth]{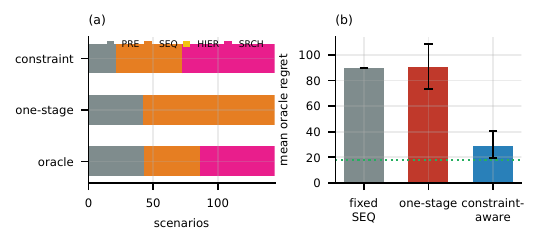}
\caption{E7 routing diagnostics. (a) Oracle membership (ties split),
one-stage selections, and secondary constraint-aware selections. (b) Mean
oracle regret with descriptive 95\% scenario-bootstrap intervals (three reused
seeds); dotted line: empirical
feature-information floor (18.1).}
\label{fgr:e7}
\end{figure}

E7 establishes a nontrivial adaptive-selection target. Search belongs to the
minimum-cost oracle set in 59 scenarios, predefined in 44, and sequential in 43;
two scenarios tie between search and predefined, while hierarchical is never
oracle. The prespecified ridge selects sequential in 102/144 scenarios and
predefined in 42 (Figure~\ref{fgr:e7}); exact-oracle selection is 31.9\%, mean
regret is 90.74 [73.76, 108.60], and value versus fixed sequential is $-0.68$
[$-4.99,3.45$].

The secondary constraint-aware analysis filters known deadline-infeasible modes
before predicting physical quality. It selects search in 72 scenarios,
sequential in 51, and predefined in 21; exact-oracle selection rises to 67.4\%,
mean regret falls to 29.00 [19.30, 40.42], and value versus fixed sequential is
$+61.06$ [39.61, 82.24]. It closes 84.8\% of the gap from fixed sequential to
an optimistic feature-cell reference. On the 72 all-feasible scenarios, however,
value versus fixed search is $-18.50$ [$-34.40,-3.88$]. The gain is feasibility-aware: a secondary parameter-free rule (search if its
nominal 5.5~s cost fits, otherwise sequential) reaches regret 28.65 and value
$+61.40$ over fixed sequential. A learning advantage is therefore not established;
within-feasible quality ranking remains unresolved by these features and estimator.

Retrospective rescoring of fixed E7 trajectories with voltage weights
$\{2000,4000,8000\}$ and curtailment weights $\{0.125,0.25,0.5\}$ changes oracle
sets in at most 5/144 scenarios. This does not test replanning under new weights.

\section{Discussion and Implications}

The benchmark resolves three dimensions often collapsed into terminal success.
\emph{Planning-strategy heterogeneity:} E2--E4 show architecture-specific
performance, robustness, and feasibility, while E7 contains predefined-,
sequential-, and search-optimal scenarios. \emph{Execution fidelity:} E5
preserves perfect mode agreement under injected, harmful objective substitution,
and E6 identifies which prosumer groups realise the directives. \emph{Adaptive
selection:} E1 reveals concentrated natural use, whereas E7 supplies a paired,
stress-held-out target for learned selectors.

The multimodel extension scopes those claims. Declaration collapse is
model-dependent: two declarers use all four strategies, but only MiniMax's
band association survives the multiplicity check; one remains concentrated
without measurable state dependence, and two are invariant. The communication interface preserves broad archetype order
for most models, but bounded mode advice can degenerate to a fixed prior. Measured
serving tails also refine the E7 design principle: deterministic structural
feasibility can still filter impossible executor types, but live deployment
requires a separate probabilistic latency margin.

The post-hoc diagnostic illustrates the value of imposing known feasibility
before quality selection; the simple rule captures essentially the same gain
without learning. The all-feasible ablation isolates physical-quality ranking as
the remaining challenge.

The study remains a controlled mechanism analysis. The physical E1--E7 bank uses one
Llama backbone, one radial topology, synthetic prosumer behaviour, linearised
power flow, and stipulated executor costs; behavioural and feeder parameters
are not calibrated to real operational data. Search shares the evaluator's
power-flow model; E3 perturbs forecasts, not model structure. The five-model
extension tests interfaces, not cross-backbone physical rankings. E1--E6 have
three to five seeds, and E7's scenario-bootstrap intervals are descriptive because
seeds recur.
The response cache statistics are preserved, but the in-memory cache itself was
not serialised. Stronger claims require full multi-backbone reruns, larger
independent seed/topology families, standard balanced and unbalanced feeders,
alternative physical evaluators, controlled latency measurements, and nonlinear
or hardware-in-the-loop validation. Both constraint-aware routing and the
simple-rule comparison are post-hoc; real-time overhead and probabilistic serving
feasibility remain unvalidated.

\section{Conclusion}

We introduced a controlled, physics-grounded benchmark for evaluating planning
strategies used by LLM agents in multi-agent cyber-physical systems. Its central
object is the planning-induced control trajectory, and its paired
counterfactuals, response-aware metrics, event-level feasibility, and external
physical referent make planning-strategy heterogeneity, execution fidelity, and adaptive
selection separately measurable.

In this single-feeder, Llama-3.3-70B study, forced search is the five-seed
baseline oracle under the specified objective, and the factorial bank contains
three oracle modes. Injected objective substitution preserves mode agreement
while increasing voltage shortfall by $2.68\times$. The prespecified ridge has
regret 90.7; secondary constraint-aware routing reduces it to 29.0, but a simple
deadline rule attains 28.7. This supports explicit feasibility handling, not a
demonstrated learning advantage. The all-feasible ablation leaves physical-quality
ranking unresolved. The five-model interface extension does not replicate
physical rankings; it shows model-dependent declarations and observed
serving-latency tails. These simulation-internal results motivate testing
constraint-aware, state-conditioned routing with probabilistic latency margins
in broader settings. The methodology can extend to other domains where an LLM
selects a planning architecture, autonomous entities react, and physics arbitrates
outcomes.

\section*{Acknowledgements}
This research was supported by the LUXEMBOURG Institute of
Science and Technology through the projects
``ADIALab-MAST'' and ``LLMs4EU''
(Grant Agreement No~101198470) and the BARCELONA
Supercomputing Center through the project ``TIFON''
(File number MIG-20232039).

\bibliographystyle{IEEEtran}
\bibliography{LLMstrategicplanning}

\end{document}